\documentstyle[12pt,aasms4]{article}

\def\etal{{et al. }}
\def\ltsim{\mathrel{\hbox{\rlap{\hbox{\lower4pt\hbox{$\sim$}}}\hbox{$<$}}}}
\def\gtsim{\mathrel{\hbox{\rlap{\hbox{\lower4pt\hbox{$\sim$}}}\hbox{$>$}}}}

\def\aa{{A\&A}}
\def\aas{{A\&AS}}
\def\aj{{AJ}}

\def\apjl{{ApJLetters}}

\def\mnras{{MNRAS}}
\def\pasp{{PASP}}

\input{psfig}

\lefthead{Beaulieu et al.}
\righthead{The Metal-rich Globular Cluster NGC6553: Observations with
WFPC2, STIS, and NICMOS}

\begin{document}

\title{The Metal-rich Globular Cluster NGC6553: Observations with 
WFPC2, STIS, and NICMOS\footnote{Based on observations with
the NASA/ESA Hubble Space Telescope, obtained at the Space Telescope
Science Institute, which is operated by the Association of Universities 
for Research in Astronomy, Inc. under NASA contract No. NAS5-26555}}

\author{Sylvie F. Beaulieu, Gerard Gilmore, 
Rebecca A.W. Elson\altaffilmark{2}, Rachel A. Johnson}

\affil{Institute of Astronomy, Madingley Road, Cambridge CB3 0HA, UK;
E-mail: beaulieu,gil,raj@ast.cam.ac.uk}

\altaffiltext{2}{Deceased, May 1999; an obituary is available in
Physics Today, September 1999, p74.}

\author{Basilio Santiago}
\affil{Instituto de Fisica, Universidade Federal Rio Grande do Sul,
91510-970 Porto Alegre, RS Brasil; E-mail: santiago@if.ufrgs.br}

\author{Steinn Sigurdsson}
\affil{Department of Astronomy \& Astrophysics, Pennsylvania State
University, University Park, PA 16802, USA; E-mail:
steinn@astro.psu.edu}

\author{and Nial Tanvir} 
\affil{Department of Physical Science, University of Hertfordshire,
College Lane, Hatfield, AL10 9AB, UK; E-mail: nrt@star.herts.ac.uk} 

\begin{abstract}

We present a HST study of the metal-rich globular cluster NGC6553
using WFPC2, NICMOS and STIS. Our primary motivation is to calibrate
the STIS broad-band LP magnitude 
against $V_{555}$ and $I_{814}$ magnitudes for
stars of known metallicity and absolute (visual) magnitude, for
application to our study of LMC globular clusters.  NGC6553 has been
shown in earlier studies to have a very unusual colour-magnitude
diagram, so we also use our data to investigate the reddening,
distance, luminosity function and structure of this cluster. 
We deduce a higher metallicity and smaller distance modulus than did
some previous studies, but emphasise that very large patchy extinction
on small angular scales prohibits accurate determination of the
parameters of this cluster. The
horizontal branch of NGC6553 in ($V,V-I$) is tilted at an angle close
to that of the reddening vector. We show that extinction does not,
however, explain the tilt, which is presumably a metallicity
effect. The colour-magnitude diagram shows an apparent second turnoff
some 1.5~magnitudes fainter than that of the cluster. We show that
this is most likely the background Galactic bulge: however, in that
case, the colour-magnitude diagram of NGC6553 is not a good match to
that of the field bulge population. The cluster is probably more
metal-rich than is the mean field bulge star.

\end{abstract}

\keywords{globular clusters: individual: NGC6553 
interstellar extinction, luminosity functions, Galactic bulge,
photometric calibrations.}

\section{Introduction}

This paper is one of a series describing the results of a large Hubble
Space Telescope (HST) project to study the formation and evolution of
rich star clusters in the Large Magellanic Cloud (LMC).  The Galactic
bulge globular cluster, NGC6553, was observed as part of this program
primarily for calibration purposes, in order to transform NICMOS and
STIS magnitudes for LMC cluster stars directly into absolute
magnitudes. Using Galactic globular clusters observed in all the HST
instruments directly ensures a calibration known to be appropriate to
the specific stellar absolute magnitude, temperature, and
metallicity. This is particularly important for determining the low
mass end of the luminosity function in the LMC clusters, one of our
primary goals. By choosing nearby Galactic clusters, we are able to
use as templates many stars with relatively well-determined
properties, with an exposure time short compared to that required in
the LMC.

NGC6553 was chosen as a suitable metal-rich template cluster on the
basis of its several detailed analyses (most recently that of Guarnieri
\etal 1998) and because of its measured metallicity of $\rm [Fe/H] =
-0.22\pm 0.05$, representative of the metallicities of the young and
intermediate age LMC clusters (cf. Olszewski \etal 1991). It later
became clear, from re-analysis of archive HST data (Feltzing \&
Gilmore 2000), that this cluster has some observational difficulties,
particularly very large and non-uniform extinction,
which make it a less than ideal calibrator. Nonetheless, it remains a
relatively metal-rich globular cluster.

Beyond calibration, our data allow us to re-investigate the
observational properties of the cluster itself. Clusters like NGC6553
are of particular interest in that, being old and metal-rich, they
are the only convenient single stellar populations representing
a typical star in giant galaxies. Metal-rich clusters are
used as a tracer of the formation and evolution of either the Galactic
bulge (in spite of their relatively large distance from the bulge),
the late evolution of the halo (in spite of their high metallicity,
and possibly inappropriate kinematics), or the early evolution of the
disk (in spite of the failure of the disk to make younger globulars).
More importantly, they extend the metallicity range over which one may
determine cluster luminosity (and mass) functions to metallicities
typical of most stars.  In this regard, the properties of the
metal-rich clusters provide clues concerning the enrichment history of
the inner Galaxy, the timescale for its formation, and the time of its
formation relative to the halo and thick disk. Their dynamically
vulnerable location also allows the effect of tidal shocking on the
structure and stellar content of the clusters to be explored.

Our WFPC2 data allow us to probe the stellar content of NGC6553 well
below its main sequence turnoff, and allow another investigation of
the cluster's distance and reddening.  Our STIS data allow us to
determine a deeper luminosity function than that previously available,
which may be compared with luminosity functions in other globular
clusters. The combination of our optical and near-IR NICMOS data
allow checks on the nature of the very 
tilted `horizontal' branch, testing the relative importances of 
extinction and astrophysical explanations, following Ortolani et
al. (1990). 

Two WFPC2 studies of NGC~6553 have been published; a detailed study by
Guarnieri \etal (1998), and an analysis as part of a larger study of
the age of the bulge by Feltzing \& Gilmore (2000).  The former is
based on F555W and F814W exposures of 200 and 100 seconds respectively
(1/3 and 1/6 the length of our exposures). The latter is based on
archive data, primarily that of Guarnieri et al. (1997), supplemented
by observations in other lines of sight. Feltzing \& Gilmore (2000)
provide a summary of the very many determinations of the metallicity,
distance and extinction towards NGC6553 which are available in the
literature. Since the precise values of these parameters are of
secondary importance here, we refer to that paper for details, and
references to the original work. The several ground-based photometric
studies are discussed and reviewed in Sagar et al. (1999).

Results derived from our WFPC2 data are described in Section 2. 
Our STIS data, together with a deep luminosity function for NGC6553
are described in Section 3. NICMOS data are described in Section 4.
Section 5 contains relations between stellar magnitudes in 
the various filters, $V_{555}$, $I_{814}$, and
$I_{LP}$. 

\subsection{The HST/LMC Globular Cluster project}

HST project GO~7307 is a study of the evolution of rich star clusters in the
LMC. A total of 95 orbits was allocated to this project in Cycle 7,
using all of the three imaging instruments in parallel: the Wide Field
and Planetary Camera 2 (WFPC2), the Space Telescope Imaging
Spectrograph (STIS) (in imaging mode), and the Near Infrared Camera
and Multi-Object Spectrometer (NICMOS) to obtain multi-wavelength
photometry in eight clusters with ages $\sim 10^7, 10^8, 10^9$ and
$10^{10}$ years. In addition, shorter exposure observations of one
Galactic globular cluster  obtained here, and observations of two
other clusters coordinated with another HST GO project (7419;
Feltzing, Gilmore \& Wyse 1999; Houdashelt, Wyse, \& Gilmore 2000),
are being used to
provide an empirical calibration of the transformation from `native'
STIS and NICMOS magnitudes to standard absolute magnitudes. The three
clusters include one very metal poor (M15), one of intermediate
metallicity (47~Tuc), and one metal-rich (NGC6553). By providing STIS
and NICMOS observations of stars of known metallicity and absolute
(visual) magnitude in these calibration clusters, we are able to
derive this calibration directly, comparing stars of similar mass and
metallicity. We will apply these calibrations to our LMC cluster
luminosity functions in future papers.  First results have been
presented by Beaulieu et al. (1999a, 1999b), Elson et al. (1999), and
Johnson et al. (1998, 2000a). The results of a pilot study of one of
the clusters in the sample are described in Elson et al. (1998a,b),
and Burleigh et al. (1999). Other recent results from this project are
presented in Santiago et al. (2000; WFPC2 luminosity functions for six
clusters), in Castro et al. (2000; WFPC2 deep CMDs of LMC field stars)
and in Johnson et al. (2000; age spreads in young LMC clusters).

A brief summary of the goals of the project is as follows: in all the
clusters we are using colour spreads among main sequence stars to look
for populations of close stellar binaries, both in the core and as a
function of radius. We will thereby determine the fraction of
primordial binaries from the youngest clusters, as well as how the
binary fraction evolves as the 
clusters age. Binaries are believed to play a crucial dynamical role
in the evolution of rich star clusters, particularly in defining the rate
of evolution of the core in old age.  Within the youngest clusters
our observations investigate the massive stellar population, and look
for age spreads.
Spreads in colour at faint magnitudes ($V\sim 25$) are used to
identify pre-main sequence stars. We will thus address the questions
of whether the high or low mass stars form sequentially or together
when a gas cloud collapses to form an approximately coeval cluster of
stars, and whether the time scale for star formation is much shorter
than, or comparable to, the dynamical crossing time. 

We are determining deep luminosity functions (LFs) in all the clusters
both in the cores and further out. In the youngest clusters this will
indicate whether there is primordial mass segregation with, for
instance, the most massive stars forming preferentially in the deepest
part of the protocluster's potential well. In the older clusters this
will enable us to trace the development of mass segregation, as the
heaviest stars sink to the centre, and investigate the formation and
evolution of the blue straggler population(s). Finally, since the stellar
populations in the young and intermediate age clusters are relatively
unaffected by dynamical selection, their LFs may be translated
straightforwardly into initial mass functions (IMFs). We can therefore
address directly the all-important question of whether these clusters
display a `universal' IMF. The as yet unproved assumption that the IMF
is universal underpins all attempts to interpret the integrated light
of galaxies at cosmological distances.

\section{NGC6553: WFPC2 Photometry}

$V$ and $I$ band data were obtained with WFPC2, with the cluster
centred on WFC Chip 3, rather than on the PC.  The scale of WFPC2 Chips
2, 3 and 4 is 0.0996 arcsec $\rm pixel^{-1}$, while the PC (Chip 1)
has a scale of 0.0455 arcsec~ $\rm pixel^{-1}$. The gain was set to
7 e-~ DN$^{-1}$. Instrumental
parameters and in-orbit characteristics of WFPC2 can be found in
Biretta (1996).  NGC6553 has a core radius $r_c=0.55$ arcmin and a
half-mass radius $r_h = 1.55$ arcmin (Harris 1996), so Chip 3 contains
somewhat more than the cluster core, while the full WFPC2 field of
view extends to slightly more than $r_h$. Table~\ref{N6553-summ}
summaries the basic properties of NGC6553. The datasets and exposure
times for the WFPC2 observations are listed in
Table~\ref{N6553-wfpc}. An image of the WFPC2 Chip 3 (core) is
shown in Figure~\ref{N6553-wfpcimage}.

The images were calibrated using the standard HST pipeline procedure,
and the long and short images were stacked separately using the IRAF
routine {\it crrej}. To identify the stars in the image we ran {\it
daofind} with a $3\sigma$ detection limit on the F814W images. There
were no spatial offsets between the F555W and F814W images, so the
F814W {\it daofind} list of objects was also used for the F555W
images. We carried out aperture photometry in F555W and F814W using an
aperture radius of 2 pixels for all chips (PC: 0.09 arcsec, WFC: 0.2
arcsec). Our experience has been that aperture photometry,
particularly on the undersampled WFC chips, produces tighter CMDs than
PSF fitting. However, PSF fitting is useful for identifying and
eliminating spurious detections via the {\it allstar} `sharpness'
parameter. This parameter is defined as the difference between the
square of the width of the object and the PSF. Therefore, we also
carried out PSF fitting using the routine {\it allstar}, for which
TinyTim PSFs were used. Note that the available STScI recipe for
correction of the systematic photometric effects resulting from CCD
CTE effects assumes 2-pixel radius aperture magnitudes.

The sharpness parameter as a function of F555W magnitude for the long
exposures is plotted in Figure~\ref{wfpc-sharp}. The results are
very similar for the short exposures and also for the F814W
image. Selection limits were typically $-0.30 < $sharpness$ <0.20$,
but varied slightly from chip to chip. We also tried PSF fitting with
a PSF constructed from stars in the image rather than from
TinyTim. This did not result in significantly tighter sharpness
distributions, and particularly for Chip 3, containing the cluster
core, the field was sufficiently crowded that isolated stars suitable
for constructing a PSF were rare.  This crowding is responsible for
the shallower limiting magnitude and the tilt in the distribution of
points in Figure~\ref{wfpc-sharp}c, though the apparent increasing
sharpness of stars in more crowded regions is somewhat non-intuitive.
We note, however, that the sharpness selection adopted is not critical
for the purposes of this present analysis.

Having removed spurious detections from both the F555W and F814W
object lists, we then returned to the aperture photometry, with a 
2-pixel aperture, and calibrated this following Baggett \etal
(1997). The calibration includes corrections for charge transfer
(in)efficiency (CTE) (Whitmore et al. 1999), geometric distortion
(Holtzman \etal 1995), aperture corrections (a correction to a 
radius of 0.5") using bright stars in the image, and zero-pointing
from Baggett \etal (1997). Colour magnitude data for 
the separate chips are shown in Figure~\ref{CMD-short1.4}a-d 
(short exposure) and Figure~\ref{CMD-long1.4}a-d (long exposure).

Figure~\ref{complVIsl}a,b show the completeness functions for V and I,
for both short (filled line) and long (dashed line) exposures.
The difference between the short and long data is largest in
the PC chip and smallest in the WFC chips, due to a better
sampling of the PSF in the PC chip, making stellar detection
more sensitive to exposure time. For the WFC chips, a stronger
limitation in image classification will arise because of pixel subsampling,
which reduces the ability to distinguish between extended sources
and close pairs of stars. The net result is that, even at the bright
end, some stars will be lost as they are merged with neighbours.
This will also happen at the faint end, but there stars will be lost
mainly because of noise which begin to dominate. The effect is most
evident in WF3 (cluster core) where the functions are the least sensitive
to exposure time because completeness is strongly limited by crowding,
not image depth. In general, the difference in the V and I exposure
times (see Table 2) seem to indicate that, where crowding is most severe,
there is not much gain in doing a long exposure.
 
The main features of the colour magnitude diagrams are discussed in
detail below. These include the blue population brightwards of the
main sequence turnoff, visible in the short exposure data, the tilted
horizontal branch (HB) and the clump below the HB, also visible in
the short exposure CMDs, and an apparent `second turnoff', an excess
of stars near $(V,V-I)=(21,2)$. The second faint red `turnoff' is
particularly prominent in chip WF4 (Figure~\ref{CMD-long1.4}d).

\subsection{WFPC2: Results and discussion}

\subsubsection{Distance, Metallicity, Reddening}

There is only a moderate concensus on the values of distance,
reddening, and metallicity for NGC6553, derived from observations
obtained from both the ground and space. From spectra of two cluster
giants, Barbuy \etal (1999) derive a metallicity $\rm [Fe/H] =
-0.55\pm0.2$. Cohen \etal (1999), from high-dispersion Keck
spectroscopy of five RHB stars deduce
$\rm [Fe/H] \approx -0.16$, revised by Carreta \etal (2001) upward to 
$\rm [Fe/H] \approx -0.06$.  Guarnieri \etal (1998) apply various methods
of estimating metallicity based on the morphology of the giant branch
both in ($K,J-K$) and ($V,V-I$) and adopt a mean value of $\rm [Fe/H] =
-0.22\pm0.05$. The overabundance of several elements relative to Fe
gives an overall abundance $Z\approx Z_\odot$ (Barbuy \etal
1999). Sagar \etal (1999) present a ground-based $(V-I)$ study of
NGC6553. From the difference in $V$ magnitude between the horizontal
branch and the brightest point of the red giant branch (RGB), they
estimate $\rm [Fe/H] = -0.1$, although this value is uncertain due to the
dramatic tilt of the HB, which itself spans 0.5 mag, and the scarcity
of stars at the brightest point of the RGB. Rather than trying to
deduce a specific value for metallicity, we show below that the colour
magnitude data favour a high metallicity, comparable to solar, in
agreement with the more recent spectroscopic results.

Guarnieri \etal (1998) derive values of reddening and distance modulus
for NGC6553 of $E(V-I)=0.95$ and $(m-M)_0=13.6$. They comment that the
largest source of uncertainty in the distance modulus is the ratio
$A_V/E(B-V)$ that is needed to convert the reddening to an absorption
(see Grebel \& Roberts 1995). Further discussion of all these studies
is provided in Feltzing \& Gilmore (2000). (Note that there is a plotting
error in the CMD of NGC 6553 in Feltzing \& Gilmore (2000), noted in
a later erratum).

Figure~\ref{CMD-short-all} shows a composite CMD for NGC6553
derived from the short exposures for all four chips. Photometric
error bars are shown. A 12 Gyr isochrone is superposed.
This is based on the Bertelli \etal (1994)
isochrones, transformed to the HST filter system by G. Worthey (1998,
private communication). The isochrone is for metallicity $\rm [Fe/H] = -0.4$,
reddening $E(B-V)=0.7$, and distance modulus $(m-M)_0=13.6$.

The same isochrone is superposed on the long exposure data in
Figure~\ref{CMD-long-all}. The general shape of the isochrones,
especially around the turnoff region, in spite of limitations
with available metal-rich isochrones, suggest that
NGC6553 requires a higher metallicity, in agreement with teh more
recent spectroscopic studies.
It is also clear that patchy extinction is affecting
any conclusions. A possibility is that NGC6553 is indeed more
metal-rich than this isochrone, and than the bulge stars. We 
note that this possibility is consistent with the measured mean
metallicity of the bulge, which peaks near $\rm [Fe/H]\approx -0.3$ (cf
Figure~2 of Wyse, Gilmore \& Franx 1997), if NGC6553 is indeed near
solar metallicity, and with enhanced [$\alpha$/Fe] values, as
suggested by some studies.

To test this further we illustrate the scale of patchy extinction in
figure~\ref{15Gy-0.0}. This shows an enlargement of the main
sequence turnoff 
region, corrected for extinction using local values. 
For this correction, we fitted for reddening in 20x20 arcsec
sub-regions of the image: the good correlation between adjacent
derived reddening values supports the validity of this approach, while
the large scatter in some regions attests to patchy extinction on even
smaller scales. As expected, the scatter around the turnoff is
significantly reduced, and the isochrones fit significantly better.
It does not, however, remove completely the tilt of the horizontal
branch. Additionally, the apparent width of the main-sequence is very
much larger than typically seen in HST globular cluster studies,
suggesting that residual extinction remain significant. 
We investigate this further in the next section.

We note that, in spite of residual uncertainty due largely to
irreducible patchy extinction, the best fit distance we derive from
this figure is $(m-M)_0=13.2$, 0.4mags shorter in distance modulus than
previous estimates. This new value, together with other parameters for
this cluster, is summarised in Table~1.

\subsubsection{Horizontal Branch Morphology}

It has been noted by many previous authors, and is very obvious in
Figure~\ref{CMD-short-all}, that the horizontal branch in NGC6553
is far from horizontal. Its range in $V$ spans $\sim$0.5 mag.
Comparison with the direction of the reddening vector shown in
Figure~\ref{CMD-short-all} shows that the orientation of the
horizontal branch follows almost exactly the reddening slope in the
CMD. Consequently, several authors have noted that much of this tilt
might be attributable to differential reddening across the face of the
cluster (cf. Barbuy \etal 1998). Others suggest that metal line
blanketing can produce a tilting of the HB at the high abundance of
this cluster (cf. Ortolani \etal 1990), so that both effects may
contribute.  We note below that the red flare of main sequence stars
near $(V,V-I)=(21,2)$, discussed further below,
may also be caused by patchy extinction. The fainter red stars are
apparent however some 1.5~magnitudes fainter than the cluster main
sequence turnoff, while what would be the corresponding horizontal
branch is apparent only $\sim~0.8$ mag fainter than the cluster HB,
complicating any single explanation of both phenomena.

In order to investigate the simplest version of the differential
extinction hypothesis, we show in figure~\ref{HB-spatial} the
spatial distribution of blue and red HB stars in Chip 3 (the cluster
core). The full sample of 176 HB stars was divided in half at the
median colour value $(V-I)=1.9$. There is a marginal tendency for the
redder stars to be found preferentially in the top right of the
distribution, suggestive of extinction effects. Nonetheless, the
red-blue separation is not perfect, implying highly structured
extinction; the dust must be patchy on scales $\ltsim$~10~arcsec.

To investigate this further we attempted to invert the extinction map,
under the null hypothesis that all the HB slope is reddening. We
assumed that the intrinsic horizontal branch is indeed horizontal in
the $(V,V-I)$ plane. We then calculated the extinction required for
each star independently, to move that star from its observed location
in the CMD to lie on a de-reddened, horizontal, horizontal
branch. This produced a set of extinction values, one per HB
star. These were then used to define a 2-D map of extinction as a
function of position across the face of the cluster. To reduce noise,
this array was then smoothed with a 4th order polynomial surface
function, generating a 2-dimensional extinction map. That map
`predicts' a value of the extinction, under this null hypothesis, at
every point on the cluster. We may then apply this map to the
observed, uncorrected HB star photometry, to see if this smoothed
extinction map has any validity: if the null hypothesis is valid, a
nearly horizontal HB should result, and in addition the scatter in the
RGB and near the main sequence turnoff should be significantly
reduced.

Figure~\ref{HB-dered} shows the outcome of this attempt to correct
for differential reddening. This figure applies to stars in the WF3 
chip (core of NGC6553), and shows the photometry before (upper panel) 
and after (middle panel) correction by the derived 
reddening map, defined by a polynomial surface fit of order~4.
The reduction in photometric scatter, and HB tilt, is at best
minimal. There is similarly only a marginal reduction of scatter in
the main-sequence turnoff data, with this reddening map.
Our attempts to derive a higher spatial resolution differential
extinction map assuming the stars in the `red flare' are really
cluster main sequence turnoff stars also failed. In figure~\ref{HB-dered}
lower panel, we show the HB data from figure~\ref{15Gy-0.0}. This allows
the reader to make 
a direct comparison of both methods used to attempt to correct for 
differential reddening.  

We thus conclude, along with Ortolani et al. (1990) and Sagar
\etal (1999), that differential reddening cannot explain all the tilt
of the horizontal branch in NGC6553, though it probably contributes.
We return to the implications for the explanation of the `red flare'
stars below.

\subsubsection{Blue straggler stars, and core structure}

\paragraph{New centre estimate}
 
The available estimation of the centre of NGC6553 is that of Harris
(1996). We have calculated an improved centre, from the distribution
of stars near the main sequence turnoff. 
This value is given in Table~\ref{N6553-wfpc}.
The new surface density profiles for stars above the turnoff, at the
turnoff, and below the turnoff, are illustrated in figure~\ref{numcts}.
The local dip in the counts at zero for the turnoff and main-sequence
stars is suggestive of some incompleteness in the very crowded central
regions, even at the corresponding bright magnitudes. As the
distribution of RGB stars shows, there is a small group of very bright
giants at the cluster centre, complicating faint photometry.

The short exposure CMDs in Figure~\ref{CMD-short-all} contain a
number of blue stars brighter than the MSTO, which may either be blue
stragglers or foreground stars belonging to the Galactic disk
population.  Most globular clusters contain blue stragglers, while
recent studies of four globular clusters show these stars to be
strongly concentrated towards the cluster centres (Ferraro \etal 1997;
Mandushev \etal 1997; Guhathakurta \etal 1998; Rey \etal 1998).
Very recently, Zoccali \etal (2001) present proper motion data showing
the candidate blue straggler stars in the central regions of the
cluster are indeed members.

To investigate the radial distribution of the blue straggler
candidates in NGC6553, we have isolated on each chip stars with
$V_{555} <19.2$ and $1.04< (V_{555}-I_{814}) < 1.74$. The numbers of
stars on each of Chips 1 to 4 are 16, 52, 121, and 62
respectively. Given that Chip 1 covers an area four times smaller than
the other chips, the surface densities are roughly equal for Chips 1,
2, and 4. Chip 3, containing the cluster core, contains twice as many
blue straggler candidates.

Figure~\ref{BSS} shows the radial surface density profile of the
blue stragglers, normalised by the surface density of main sequence
stars with $19.6 \le V \le 20.1$.  This magnitude range was chosen as both
the stellar mass and the number count completeness for the main
sequence stars and blue stragglers is similar, so this ratio is
robust.  The profile is centered on WF3 (471,478).

The true surface density of blue stragglers peaks at the origin in
this co-ordinate system. The main sequence star counts plot is
very flat (Figure~\ref{numcts}) and has, if anything, a slight dip
at the coordinate centre (471,478), which may be due to
crowding-induced incompleteness at the origin. If so,
this may slightly overstate the ratio of blue stragglers to
main sequence stars at the origin, but as a 10-20\% uncertainty.  Small
number statistics are important here.  There are only 7 blue straggler
stars in the innermost bin; at the first maximum spike in the plot
there are only 6 blue stragglers, though the second spike is robust,
with 12 blue stragglers in the annulus beyong 3 core radii. That is, there
are relative maxima in the radial surface density distribution of blue
stragglers at both small and large radii.

The blue straggler radial distribution in another cluster, M3 (Ferraro
et al. 1997) shows a similar anomalous distribution, and has been
modelled by Sigurdsson et al. (1994) under the assumption that the
main mechanism for producing blue stragglers is collisions between
primordial binaries and single stars. In this model, recoil during
collisions displaces blue stragglers from the core on formation, and
produces an initial $r^{-2}$ radial distribution out to some radius
determined by the depth of the cluster potential. Blue stragglers have
higher mean mass than the typical main sequence star in the cluster,
so dynamical friction then brings the blue stragglers back into the
cluster core. The blue straggler distribution is thus enhanced in the
core, and depleted out to some critical radius $r_b$ where the
dynamical friction time scale becomes comparable to the blue straggler
lifetime. Assuming a dispersion $\sigma \sim 7~ {\rm km\, s^{-1}}$
(Cohen et al. 1999) and a core density, $\rho_0 = 10^5 M_{\odot}\,
{\rm pc^{-3}}$, we expect substantial depletion of blue stragglers. 

The
density distribution of blue stragglers rises inside the core radius
as the blue stragglers have a higher mean mass than the mean mass of
the core. The cluster core radius here is 30'', so the blue stragglers
would be expected to form a dynamical sub--core with
radius of about 20''. The density of the blue stragglers should rise
as $\sim r^{-2}$ between 20'' and 30'' and thus the surface density
should $\propto r^{-1}$. Outside the core, going from about a core
radius to about the half--mass radius, the density distribution of
blue stragglers should be steeper than $r^{-2}$ in the region where
dynamical friction efficiently transports the stars to the core, and
then the density flattens out again near the (projected) radius where
the the dynamical friction time scale become comparable to the blue
straggler lifetime. Outside this radius we predict the surface density
distribution of blue stragglers should decline again, but less steeply
than the density of the main sequence stars.

\subsubsection{Red Giant Branch Luminosity Function}

The luminosity function of the red giant branch is a test of stellar
evolutionary models, and in particular provides a consistency test of
models of old metal-rich stars. Figure~\ref{RGB} shows an
enlargement of the red giant branch region of the short exposure CMD
for all four chips and the corresponding luminosity function.  The
horizontal branch shows up prominently, peaking at $V_{555}=16.7$. A
second peak is clearly visible $\sim 0.9$ magnitudes below the HB, and
may even be tilted parallel to the horizontal branch, indicative of
patchy extinction as an explanation. The second peak has
been discussed, for example, by Lanteri Cravet et al. (1997), and by
Sagar \etal (1999) and attributed to a phase of stellar evolution
where the star loops in its ascent of the RGB. Deferring to those
analyses, we note only that the number of stars in this `RGB bump'
remains poorly determined. Figure~\ref{RGB} suggests some 30\%,
significantly larger than the 19\% of Lanteri Cravet et al., and the
17\% of Sagar et al.

Another suggestion is that the `RGB bump' may be confused by
superposition on the cluster RGB of background field HB stars,
associated with the second turnoff discussed below. In this case, the
slope cannot be attributed to common differential extinction, since
the background would have to suffer less extinction than the
foreground cluster. A further problem is that the
`second bump' has the same colour, $(V-I \approx 2)$, as does the
`second turnoff': in general, one expects a metal-rich old population
to have a horizontal branch which is redder than the main-sequence
turnoff colour, as shown by the isochrone in
Figure~\ref{CMD-short-all}. A stellar evolutionary explanation of
the observed RGB bump seems the most likely.

\subsubsection{The Red Flare}

The colour-magnitude data shown here, and in earlier published studies
of this cluster, indicate a substantially broader main-sequence width
than is typical for HST data, with the scatter in colour increasing
substantially fainter than $V$=21. Indeed, the longer exposure
colour-magnitude data show an asymmetric colour distribution about the
main sequence ridge line, especially visible as a redward extension of
the data near $(V,V-I)=(21,2)$ in chips 1 and 4
(Figures~\ref{CMD-long-all}a,d). This may be equally described as
an apparent ``second turnoff", implying a background population, or as
a `red flare', implying a part of the NGC6553 stellar population which
is substantially more reddened than the average.

Can the excess be due to extinction? Casual inspection of
Figure~\ref{CMD-long-all} shows that this faint, red group of
stars lies in the CMD in a direction close to that of the reddening
vector, relative to the cluster turnoff. Could these faint red stars
then be cluster members, obscured by up to an additional
$E(B-V)\approx 0.7$ magnitudes? Patchy extinction is also a plausible
explanation for the morphology of the horizontal branch in this
cluster. As described and investigated above, the slope of the HB is
very similar to the slope of the reddening vector. We note that, as is
also shown above for the HB, since this extra red population is not
spatially highly segregated from stars with apparently much smaller
reddening, this equivalently implies either extinction {\sl inside}
the cluster itself, or patchy extinction on very small spatial scales.
This possibility is investigated in detail above, in the discussion of
the horizontal branch morphology, and shown not to be viable.

We may test directly if this feature corresponds to a background
population, such as the turnoff of the inner Galactic bulge.  The most
important conclusion is that it is not possible to fit the fainter
turnoff with the same isochrone as for NGC6553, simply shifted to a
larger distance.  A larger distance modulus and a smaller value of
reddening are required. If the bulk of the bulge stars are at a
distance of 8.0 kpc, then the reddening required to fit the turnoff is
$E(B-V)=0.87$. This fit is illustrated by the dotted lines in
Figure~\ref{CMD-short-all} and Figure~\ref{CMD-long-all}. The
required extinction in this case is less than that for the cluster,
although that is some 3 kpc closer to us. Might there be a patch of
extinction covering just the cluster, with lower extinction adjacent?
This is difficult to arrange, as the background bulge stars, if that
is what they are, are seen through the cluster, rather than off to one
side of it.  

In principle, a bluer isochrone, that is a more
metal-poor population, at the distance of the bulge, and with
extinction comparable to that in the foreground, is feasible. Thus,
the `red flare' is consistent with being the main sequence turnoff of
the background bulge field stars, at a distance of 8~kpc, but only if
the bulge field population is more metal-poor than is NGC6553. There
are some stars, visible in the short exposure CMDs in
Figure~\ref{CMD-short1.4}a-d, which might be considered members
of a corresponding giant branch, but too few to help in the present
analysis. They may support a future comparative spectroscopic
abundance analysis with the members of NGC6553.

\subsubsection{Conclusions from the WFPC2 data}

We have tested the hypothesis that all the complexities of
the observed optical colour-magnitude data, especially the tilted
horizontal branch and the existence of the red flare, are an artefact
of very patchy extinction. We exclude this hypothesis above. This then
requires an astrophysical explanation for the slope of the horizontal
branch, with high cluster metallicity being the most obvious. This is
also consistent with the observed cluster RGB bump in the luminosity
function, as noted by earlier workers. What then is the red flare? The
best available explanation is that it is the background bulge turnoff
field stars viewed through the cluster. In this case, the bulge field
stars are not consistent with the NGC6553 CMD simply shifted to larger
distance: the bulge cannot have the same metallicity as does the
cluster, but must be more metal poor. This conclusion is consistent
with published spectroscopic data for the bulge if the cluster
abundance is indeed near solar. That is, NGC6553 is not a simple
template match to the mean Galactic bulge star, being probably more
metal-rich.

Nonetheless, all conclusions on the age, distance,
and metallicity of NGC6553, and the nature and/or reality of the `red
flare'  which depend sensitively on photometry are
clearly suspect, given the extreme and patchy extinction.

\section{NGC6553: STIS Photometry}

Our STIS image of NGC6553 is centered at
$\rm{RA}(J2000)=18^{\rm{h}}~09^{\rm{m}}~11^{\rm{s}}$,
$\rm{Dec.}(J2000)=-25^\circ~54'~33\arcsec$, about 1 arcmin from the
cluster centre, and overlaps with the PC images described above.
Three images were taken with exposure times 30, 300, and 2046 sec. The
datasets are listed in Table~\ref{STIS-data}. An image of the
field is shown in Figure~\ref{STIS-image}. In imaging mode, the
STIS detector has a size of $28 \times 51$ arcsec and a scale of 0.05
arcsec $\rm pixel^{-1}$. We used the STIS longpass filter ($I_{LP}$) which
essentially spans Johnson-Cousins $V, R$ and $I$. We discuss
photometric transformations of this passband further below, but show
that the STIS bandpass is very similar to the $I$-band for globular
cluster stars. Instrumental parameters and a summary of in-orbit
performance of STIS can be found in Sahu (1999).

The exposures were calibrated using the standard pipeline ({\it
calstis}), and stacked using the task {\it drizzle}. The resulting
summed image was run through our own photometry pipeline, which
consists of identifying stars, fitting them with PSFs constructed from
isolated stars in each image, and discarding objects with anomalous
values of the {\it allstar} sharpness parameter, as for the WFPC2
data. The
sharpness parameter, as a function of magnitude, is plotted in
Figure~\ref{STIS-sharp}, which may be compared with the
corresponding figure for the WFPC2 data in
Figure~\ref{wfpc-sharp}. Our selection limits of $-0.10 <
$sharpness$<0.15$ are indicated. There is a clear separation between
point sources and detections which are slightly resolved (above the
top line), which fall mainly along diffraction spikes and in the wings
of saturated stars. Our final magnitudes are derived from 2-pixel
radius aperture photometry. We make no correction for STIS CTE, since
there is no significant correction at GAIN=4, appropriate for our data
(Gilliland, Goudfrooij \& Kimble 1999).
 
Our pipeline also determines completeness as a function of magnitude
from a grid of 209 artifical stars added to the stacked image, and
recovered along with the real stars. Completeness as a function of
magnitude is shown in Figure~\ref{STIS-complete}.

\subsection{A STIS luminosity function for NGC6553}

Figure~\ref{STIS-LF} shows the STIS luminosity function for NGC
6553, both raw and corrected for incompleteness. The corresponding
luminosity function for 47~Tuc is overplotted for comparison. We have
also converted the STIS measured instrumental magnitude to an absolute
magnitude in the STIS $I_{LP}$ passband. We do this converting from
our utilised aperture radius of 2 pixels (0.1 arcsec), with our
measured aperture
correction of $-0.5$~mag to an aperture correction radius at 0.5
arcsec, a photometric zeropoint in AB magnitudes of $\rm ZP_{AB} =
25.28$ (Gardner et al. 2000), a reddening of $E(B-V)=0.7$ and an
absolute distance modulus of $(m-M)_0=13.6$. The reddening and
distance information are from Guarnieri \etal (1998). The value of
$E(B-V)$ has been transformed to an absorption in the STIS $I_{LP}$ passband
by $A_{I_{LP}} = 2.505~ E(B-V)$ by integration across the passband.
We also transform the STIS instrumental magnitude into an absolute
magnitude in the $I_{814}-$passband, using our own STIS-LP versus
WFPC2 relation derived below (cf. \S 5.1). This facilitates comparison
with other published HST cluster luminosity functions and isochrones.

In Figure~\ref{STIS-LF} we show also a comparison of seven
available HST globular cluster luminosity functions, including now our
transformed data for NGC6553, and other data from the literature
(Elson et al. 1995; Santiago et al. 1996; Piotto et al. 1997; de
Marchi 1999).  While the luminosity functions agree well at brighter
magnitudes, the faint ends differ markedly. Such differences have been
noted before, and have been attributed to metallicity effects,
evaporation of low mass stars, relative binary fractions, and
stripping of low mass stars by tidal shocking (cf. von Hippel et al.
1996; Piotto et al. 1997; Piotto \& Zoccali 1999; Elson et al. 1999;
Paresce \& de Marchi 2000). All these effects must apply at some
level, and will have different levels of importance in different
clusters. The metallicity-dependence of the mass-luminosity relation
will provide a systematic change in luminosity function shape with
abundance (von Hippel et al. 1996). Mass differences, and differences
in core parameters, will ensure that different clusters are in a
different state of dynamical evolution towards eventual
evaporation. Location in the Galaxy will induce different tidal
stripping effects. 

While full discussion of the relative importance of
these effects is beyond the present paper, we do recall the result of
Elson et al. (1999), who showed there is a rather good correlation
between the shape of the cluster luminosity function, as shown in
Figure~\ref{STIS-LF}, and the distance of the cluster from the
Galactic Plane. This latter quantity of course correlates strongly
with the external tidal field. The final panel of
Figure~\ref{STIS-LF} is an updated version of the corresponding
figure from Elson et al. (1999).  Further analyses of these important
and interesting effects are underway, using the latest GRAPE N-body
calculations and our HST data (Aarseth 1999).

\section{ NGC~6553:  NICMOS Photometry}

Our NICMOS NIC2 field is located at 
$\rm{RA}(J2000)=18^{\rm{h}}~09^{\rm{m}}~11^{\rm{s}}$,
$\rm{Dec.}(J2000)=-25^\circ~54'~33\arcsec$. These coordinates overlap
the region covered by the PC. The scale of the NIC2
camera is 0.0757 arcsec $\rm pixel^{-1}$, giving a field of view
of 19.4 arcsec on a side. We have obtained two exposures of 160 and
514 sec in F110W ($\approx J$) and two exposures of 576 and 1026 sec in
F160W ($\approx H$) in MULTIACCUM mode. Instrumental parameters and 
in-orbit characteristics of NICMOS can be found in Dickinson 
(1999). Details of the  
datasets can be found in Table~\ref{NIC-data}. An image of the
NICMOS F110W field is shown in Figure~\ref{NIC-image}. 

The images were recalibrated using the standard pipeline {\it
calnica}, and stacked using the task {\it mscombine}. Aperture photometry
was performed following the proposed NICMOS recipe and photometric
calibration cookbook (www.stsci.edu/instruments/nicmos/ and
nicmos data handbook). We used the IRAF/DAOPHOT task PHOT with an
aperture radius of 0.15 arcsec and an aperture correction for a radius
of 0.5 arcsec. The aperture corrections were obtained using 7 stars
on each images. The mean value of the aperture corrections with
their associated $\sigma$ are: $J_{110}:(1.505,0.032)$ and
$H_{160}:(1.678,0.032)$. A correction to a nominal infinite aperture was 
performed by multiplying the measured countrate of each star by
1.15. We then converted the countrates into flux, and then into
magnitudes ($J_{110}$ and $H_{160}$) using the photometric keywords
PHOTFNU and ZPVEGA obtained from the NICMOS Handbook (Dickinson 1999).
Table~\ref{NIC-words} lists the values used in this paper.
Care was taken in order to mask out the coronographic hole
region (radius $\approx 7$ pix at the position near (184,44)) as it 
can be mis-identified for a star in the IRAF/DAOPHOT routines.

Figure~\ref{nic-cmds} shows the near-IR NICMOS colour-magnitude
diagrams,  ($J,(J-H)$) and ($H,(J-H)$) derived from the F110W and F160W
images. The width of the main sequence near the turn-off in these
diagrams is substantially less than in the optical diagrams shown
above, confirming that non-uniform extinction is indeed a factor
constributing to the appearance of the colour-magnitude diagrams for
this cluster. This is most clearly illustrated by comparison of
figure~\ref{nic-cmds} with figure~\ref{nic-cmds2}, showing the
optical-infrared colour magnitude diagram. The substantially increased
scatter in the width of the main sequence is apparent. These
two-colour relations remain  a valuable resource for comparison with
metal-rich near-IR isochrones, in spite of this complexity.

For completeness, we also show here figure~\ref{nic-2col}, the
optical-infrared two-colour diagram. The points in this figure clearly
do not scatter along the expected narrow single-age, single-abundance,
single-extinction locus. Rather, a substantial minority of stars is
seen to be offset, along the direction of the reddening vector, by an
amount of some 30\% of the total extinction. Note that this variation
is seen inside the very small NIC2 field of view. This confirms
earlier suspicions that the colour magnitude diagram, luminosity
function, etc, of NGC6553 is significantly affected by patchy
extinction. In agreement with our analysis of the slope of the
horizontal branch, however, not all stars, and not all features of the
colour-magnitude diagram, can be explained in this way.

\section{STIS-WFPC2 Calibrations}

Our primary goal in observing NGC6553 as part of this project was to
provide a calibration to convert STIS  magnitudes to the
much better studied WFPC2 system, and hence to convert STIS apparent
magnitudes for stars in the LMC to stellar absolute luminosities in a
standard system. With this in mind, NGC6553 was chosen specifically
because its metallicity is similar to those of young and intermediate
age LMC clusters.

\subsection{The relationship between STIS $I_{LP}$ and WFPC2 magnitudes}

Our STIS image of NGC6553 intentionally overlaps the WFPC2 PC field, 
so that we can compare directly
the magnitudes of the same set of stars derived from the two instruments.
Figure~\ref{STIS-WFPC} show a comparison of STIS-LP with the
long exposure F555W  and F814W magnitudes from 1646 and 1623 stars
common to both samples respectively, and the resulting colour
dependence of the STIS {\sl vs} WFPC2 filter systems.

The relationship between the STIS $I_{LP}$ and the WFPC2 $V_{555}$ and
$I_{814}$ magnitude systems must be essentially linear, since all
three are magnitudes, with a zero point offset, and a colour term to
be determined. Since the STIS $I_{LP}$ photometric system is very
broad, we allow for this colour term to be non-linear, and we expect
it to be a function not only of stellar absolute magnitude, but also
of stellar metallicity. It is for this reason that we have studied
three calibration clusters, with a range of metallicity. In this paper
we consider NGC6553, with $\rm [Fe/H]\approx-0.2$. In a companion paper
(Houdashelt, Wyse \& Gilmore 2001) we derive corresponding
relationships for M15 ($\rm [Fe/H]\approx-2.2$) and for 47~Tuc
($\rm [Fe/H]\approx-0.7$), and consider the metallicity-dependence of
the resulting calibrations.

We expect to define relations of the following form:
\begin{equation}
V_{555} = I_{LP} + b_1(V_{555}-I_{814}) + c_1(V_{555}-I_{814})^{2}+ ZP_V
\end{equation}
\begin{equation}
I_{814} = I_{LP} + b_2(V_{555}-I_{814}) + c_2(V_{555}-I_{814})^{2}+ ZP_I
\end{equation}

The coefficients in these fits were determined using an iterative
$2-\sigma$ clipping algorithm, and are
presented in Table~\ref{phot-coeff}, while the data and the fits
to the data are shown in figure~\ref{STIS-WFPC}.

\section{Conclusions}

The metal-rich globular cluster NGC~6553 has long been known to have a
complex colour-magnitude diagram. We have obtained HST optical (WFPC2 and
STIS) and near-infrared (NICMOS) observations of NGC~6553
to calibrate photometric conversions between the various HST
passbands. In addition to that calibration, which we provide, we
utilise the data to study the cluster. The optical colour magnitude
diagram shows a curving RGB, indicative of high metallicity, a tilted
`horizontal branch', a very broad main sequence, and a `red flare' of
stars to the red of the main sequence fainter than the turnoff. We use
our combined data to support earlier suggestions that variable
extinction explains some, but not all, of these peculiarities. The
`horizontal branch' is indeed tilted, at least in part, by an
astrophysical cause. The width of the main sequence is caused, in
large part, by patchy extinction. The `red flare' is however most consistent
with being the background bulge field stars, seen through the cluster.
In this case, NGC~6553 is more metal-rich than is the mean field bulge
star. 

All conclusions relevant to this cluster, including ours, are
seriously compromised by the large and patchy extinction, allowing the
possibility of significant residual systematic errors, and making
meaningless any formal `error bars' on derived quantities. NGC6553 is
affected by variable extinction of amplitude some 30\% of the total
line of sight extinction towards the cluster. It is this variable
extinction which complicates analysis of the photometry for this
cluster.  Nonetheless, our best efforts to reproduce the photometric
data, while remaining imperfect, suggest an age of 13Gyr, a solar
abundance, and a distance modulus 0.4mags smaller than that of
previous studies.  The extinction varies markedly on small scales,
invalidating use of a single extinction in photometric studies
supporting spectroscopic abundance determinations. Our isochrone study
supports a near-solar abundance.  This high metallicity is consistent
with the general shape of the colour-magnitude diagram.


\clearpage


\begin{figure}
\caption{A visual image of NGC6553. This image is from WFPC2, and
shows Chip 3, centred on the cluster core. The cluster appears
essentially smooth and regular, even though patchy extinction may
explain some of the properties of its colour-magnitude diagram.
The arrow indicates north, the line east. SEE FIGURE 1 (JPG) 
\label{N6553-wfpcimage}}
\end{figure}

\clearpage

\begin{figure}
\caption{ Sharpness parameter from PSF fitting (with TinyTim PSFs)
as a function of magnitude, for the long exposures taken with the
F555W filter of NGC6553 for the four WFPC2 chips. The adopted
range of sharpness for real stars is between the dashed lines. Points
falling above the top line are generally associated with diffraction
spikes, while those lying below the bottom line are generally warm
pixels or residual cosmic rays. The bottom left panel is for the
cluster core (Chip 3). Here, crowding accounts for the fact that the
points do not go as deep, and for the trend in sharpness with
magnitude. SEE FIGURE 2 (JPG) \label{wfpc-sharp}}
\end{figure}

\clearpage

\begin{figure} 
\caption{Colour magnitude data for the short exposures of NGC
6553. The length of the diagonal dash indicates the extinction in
this field, with the slope that of the reddening vector. The
similarity of the slopes of the horizontal branch and of the
reddening vector are evident.  Brighter stars than those shown
are saturated. The giant branch of the background bulge population
is most clearly visible redwards of the NGC6553 giant branch in
Chips 2 and 4. SEE FIGURE 3 (JPG) \label{CMD-short1.4}}
\end{figure}

\clearpage

\begin{figure}
\caption{Colour magnitude data for the long exposures of NGC 6553.
The diagonal line indicates the extinction, as in
Figure~\ref{CMD-short1.4}. Brighter stars than those shown are
saturated. Note the presence of a red flare of stars, near
$(V,V-I)=(21,2)$, most clearly visible in Chip 4. We suggest this
is the main-sequence turnoff of the background bulge population.
SEE FIGURE 4 (JPG) \label{CMD-long1.4}} 
\end{figure}

\clearpage

\begin{figure}
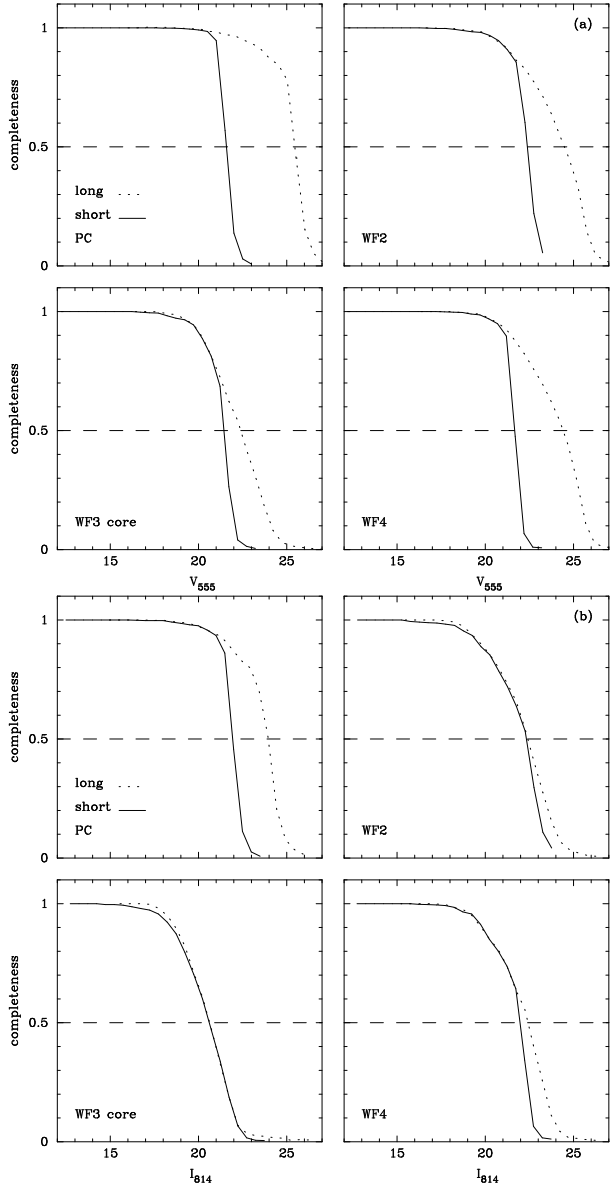

\psfig{figure=BeaulieuSF.f5a.ps,width=8.0cm}
\psfig{figure=BeaulieuSF.f5b.ps,width=8.0cm}
\caption{Completeness functions of NGC 6553 for $V_{555}$ (a) and
$I_{814}$ (b) with short (filled line) and long (dashed line) exposures.
The cluster's core is on the WF3 chip. The $V_{555}$ and $I_{814}$
exposure times can be found in Table 2. 
\label{complVIsl}}
\end{figure}

\clearpage

\begin{figure}
\caption{Combined colour magnitude data for the short
exposure observations of NGC6553 from Chips 1,2,3 and 4.
A 12 Gyr isochrone for $\rm [Fe/H]=-0.4$, reddening $E(B-V)=0.7$,
and distance modulus $(m-M)_0=13.6$, values from the literature,
is superposed. This isochrone is
not a good fit to the main sequence slope. The dashed isochrone
represents the background bulge stars and has $\rm [Fe/H]=-0.4$,
$E(B-V)=0.87$, and distance 8 kpc.
The diagonal line in the top right of the figure
indicates the amount by which
the isochrone has been offset to allow for reddening.
SEE FIGURE 6 (JPG) \label{CMD-short-all}}
\end{figure}

\clearpage

\begin{figure}[h]
\caption{Combined colour magnitude data
for the long exposure observations of NGC6553 from Chips
1 and 4. A 12 Gyr isochrone for $\rm [Fe/H]=-0.4$, reddening
$E(B-V)=0.7$
and distance modulus  $(m-M)_0=13.6$ is superposed, but does not
represent the main sequence slope well, as in figure~5. The
dashed curve is
a 12 Gyr isochrone with the same metallicity for a distance of
8.0 kpc, with reddening $E(B-V)=0.87$, expected for a background
bulge population. SEE FIGURE 7 (JPG) \label{CMD-long-all}}
\end{figure}

\clearpage

\begin{figure}
\caption{Colour-magnitude data corrected for reddening in 20x20 arcsec
regions. The overlaid isochrones are for $\rm [Fe/H]=0.0$ and -0.4, and
age 13.18 Gyr. The fit around the turnoff and subgiant branch is clearly 
improved relative to use of a single extinction value. The deduced
distance modulus is 0.4mags smaller than that derived in previous 
studies. SEE FIGURE 8 (JPG) \label{15Gy-0.0}}
\end{figure}

\clearpage

\begin{figure}
\psfig{figure=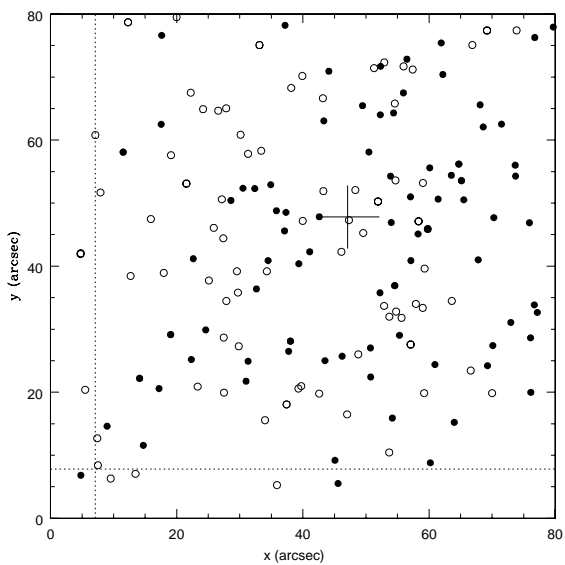,width=8.0cm} 
\caption{To
test patchy extinction as an explanation of the tilt of the
horizontal branch, we show the spatial distributions of the 176
horizontal stars in the cluster core divided into two groups: those
bluer than $V-I=1.9$ (open circles) and those redder (filled
circles). The dotted area is centred on the cluster centre which is
indicated with the cross at (471,478).  Some blue-red segregation is
apparent, indicating that differential extinction has at least
contributed to the morphology of the horizontal branch.
However, the lack of a clear segregation suggests that another effect
is also relevant. \label{HB-spatial}}
\end{figure}

\clearpage

\begin{figure} 
\psfig{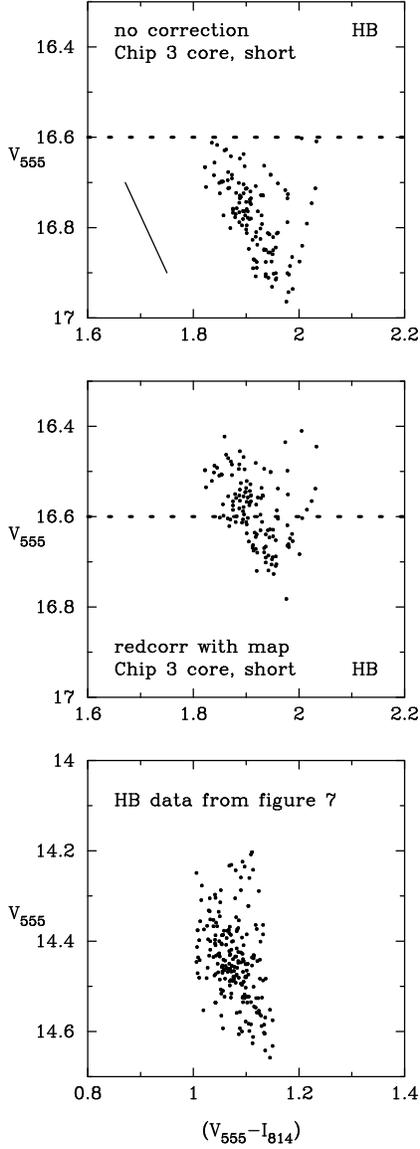} 
\caption{The upper panel is an enlargement of the colour magnitude
data for the horizontal branch stars from Figure~3. The diagonal
line is the reddening vector, which is seen to have a similar,
but not identical, slope to that of the `horizontal' branch.
The middle panel shows the same stars, after derivation and application 
of a 2-dimensional extinction map, derived assuming the horizontal
branch is indeed horizontal, and then smoothed in spatial distribution.
The dashed line indicates the assumed location of this
horizontal branch.  This assumption is clearly not supported by
the extinction-corrected data, showing that much of the morphology
of the horizontal branch is intrinsic to the stars, and not an
artefact of variable extinction on spatial scales of several
arcseconds or larger. Finally, the lower panel shows the HB data
from figure 7, illustrating the residual astrophysical contribution
to the tilt of the HB. \label{HB-dered}}
\end{figure}

\clearpage

\begin{figure}
\psfig{figure=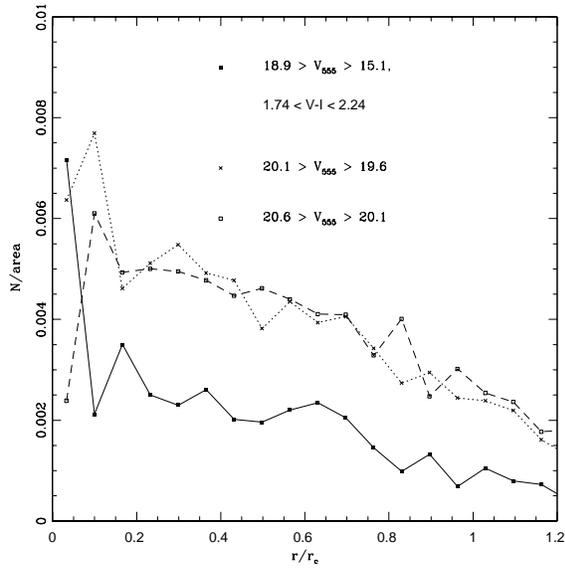,width=8.0cm}
\caption{Radial surface density profiles for the three different
luminosity groups
of stars indicated. The brightest group is the RGB, the second group
stars near the main sequence turnoff, and the third group stars just
below the turnoff. The core radius is $\sim~30$ arcsec.
\label{numcts}}
\end{figure}

\clearpage

\begin{figure}
\psfig{figure=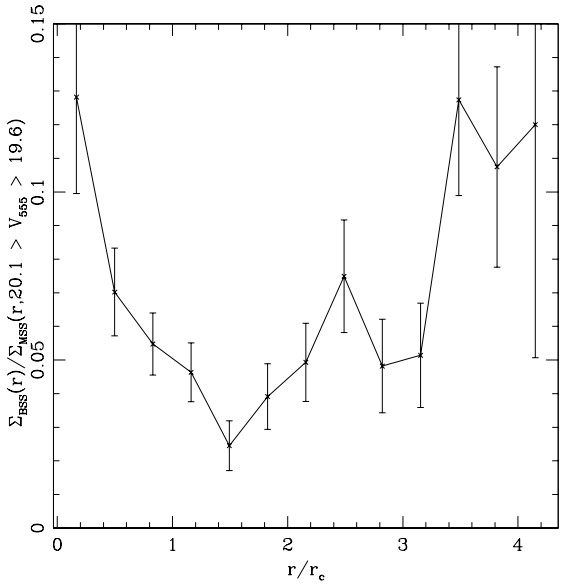,width=8.0cm}
\caption{The ratio of the surface density of blue straggler to main
sequence turnoff stars in NGC6553, in the apparent magnitude range
indicated in the caption. There is a relative excess of blue
stragglers at both small and large distances from the cluster centre.
\label{BSS}}
\end{figure}

\clearpage

\begin{figure*}
\centerline{
\psfig{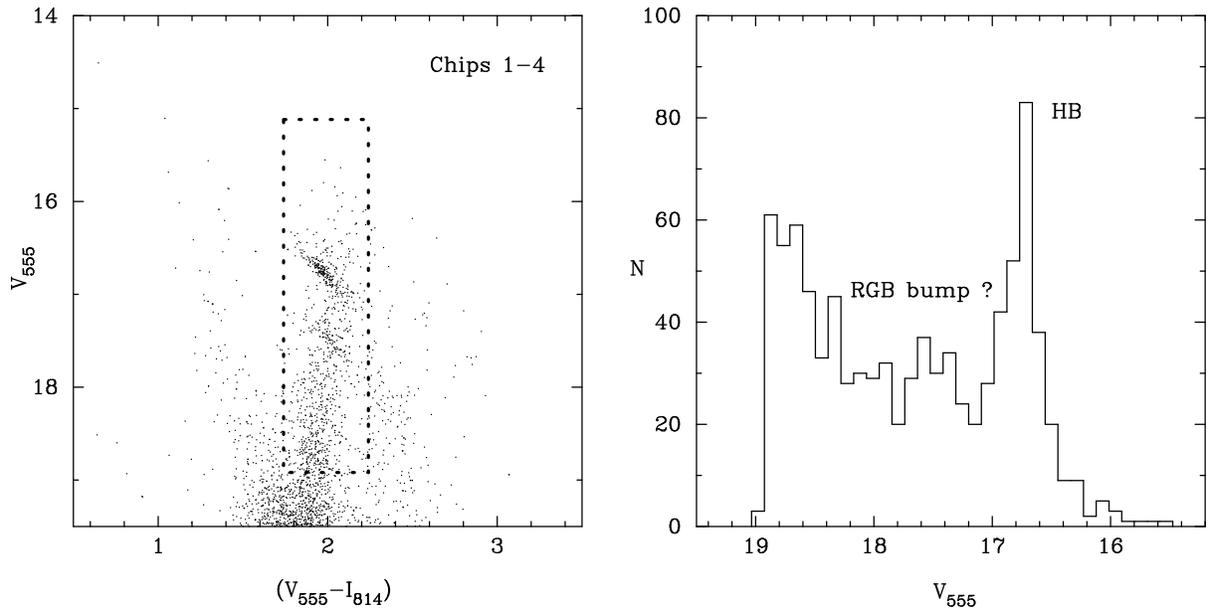}}
\caption{An enlargement of the red giant branch region of the colour
magnitude diagram from Figure~3 (left panel). The sloping horizontal
branch is apparent, as is the concentration of stars some one
magnitude fainter. The area outlined by the box is that used for
derivation of the giant branch luminosity function (right panel).
The luminosity function maxima due to the horizontal branch and `RGB
bump' are clearly visible. \label{RGB}}
\end{figure*}

\clearpage

\begin{figure}
\caption{STIS image of NGC6553. The image has size $29 \times 58$
arcsec. The faintest stars visible have magnitudes $M_I\sim 24$. The
arrow indicates north, the line east. SEE FIGURE 14 (JPG) 
\label{STIS-image}}
\end{figure}

\clearpage

\begin{figure}
\psfig{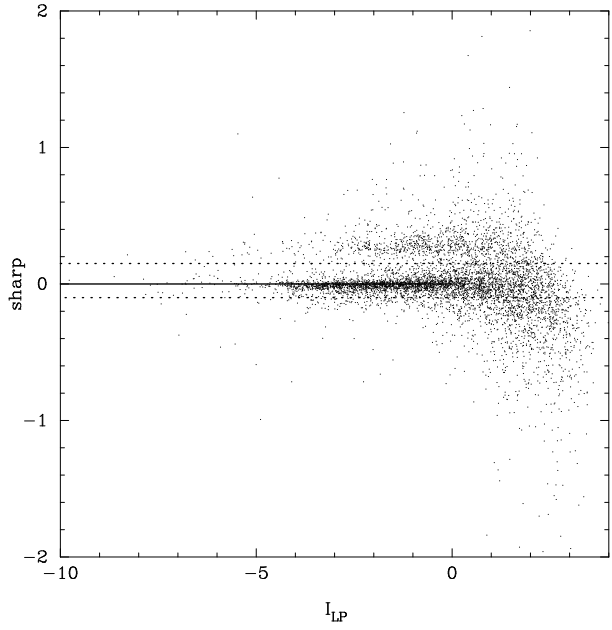}
\caption{ Sharpness parameter from PSF fitting plotted as a function
of magntiude for STIS photometry of NGC6553. Points selected as stars
fall between the dotted lines. Points falling above the top line are
generally
associated with diffractions spikes, while those lying below the
bottom line are generally warm pixels. \label{STIS-sharp}}
\end{figure}

\clearpage

\begin{figure}
\psfig{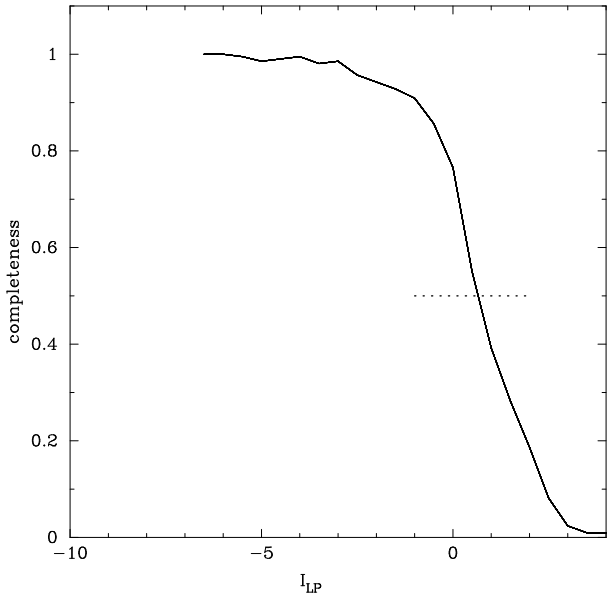}
\caption{Completeness as a function of STIS magnitude for NGC6553.
The horizontal line indicates the 50\%  completeness limit.
\label{STIS-complete}}
\end{figure}

\clearpage

\begin{figure*}
\centerline{
\psfig{figure=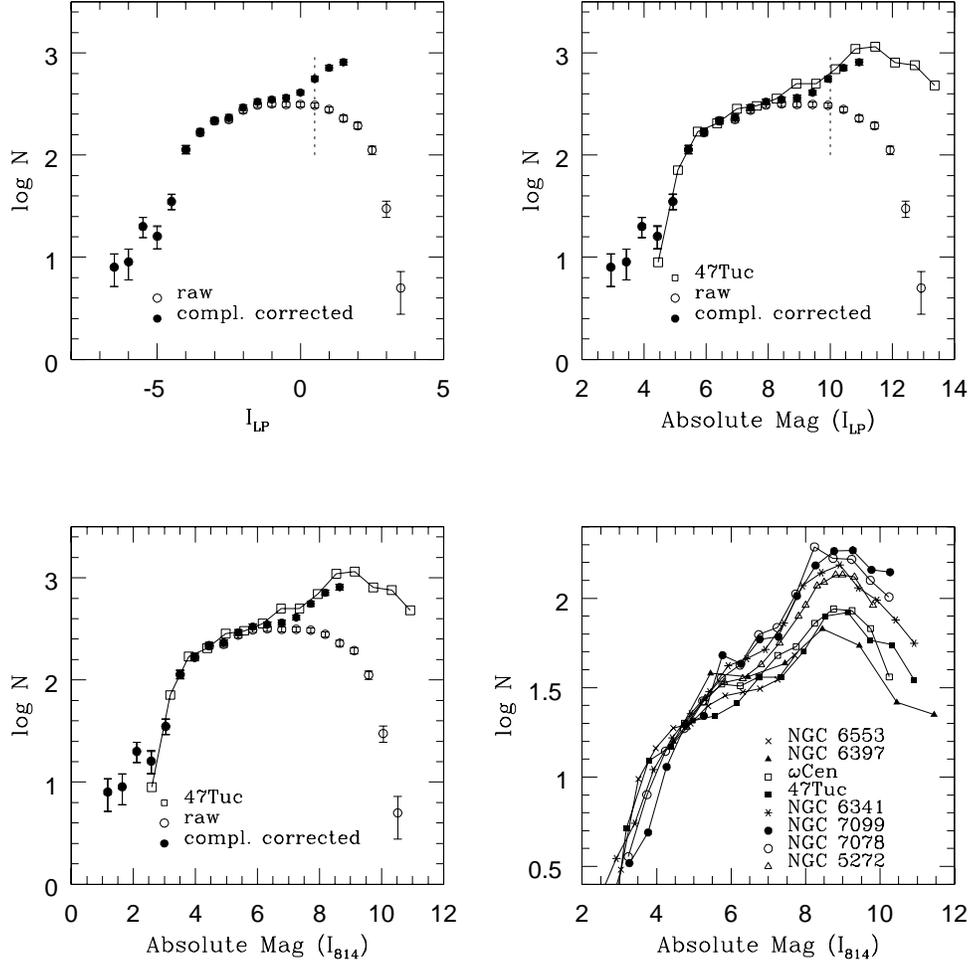,width=14.0cm}}
\caption{STIS and $I$-band luminosity functions,
for a field $\sim 1$ arcmin from the centre of NGC6553. The top left
panel shows the raw and completeness corrected luminosity function in
the STIS instrumental system. The top right shows the same function
converted to absolute STIS magnitude, after reddening and distance 
modulus correction. In both of these the vertical dashed line
indicates the 50\%  completeness limit. The bottom left panel shows
the luminosity function for NGC6553, after conversion of the STIS
absolute magnitudes into absolute $I_{814}$ magnitudes, using the
transformation of this paper. The luminosity function of 47~Tuc is
also shown, for comparison. The bottom right panel shows absolute
$I_{814}$ luminosity functions for seven globular clusters, including
NGC6553. \label{STIS-LF}}
\end{figure*}

\clearpage

\begin{figure}
\caption{NICMOS NIC2 $J$-band (F110W) image of NGC6553. The arrow 
indicates north, the line east, and the spatial scale at the cluster 
is indicated by the scale bar. SEE FIGURE 18 (JPG) \label{NIC-image}}
\end{figure}

\clearpage

\begin{figure}
\psfig{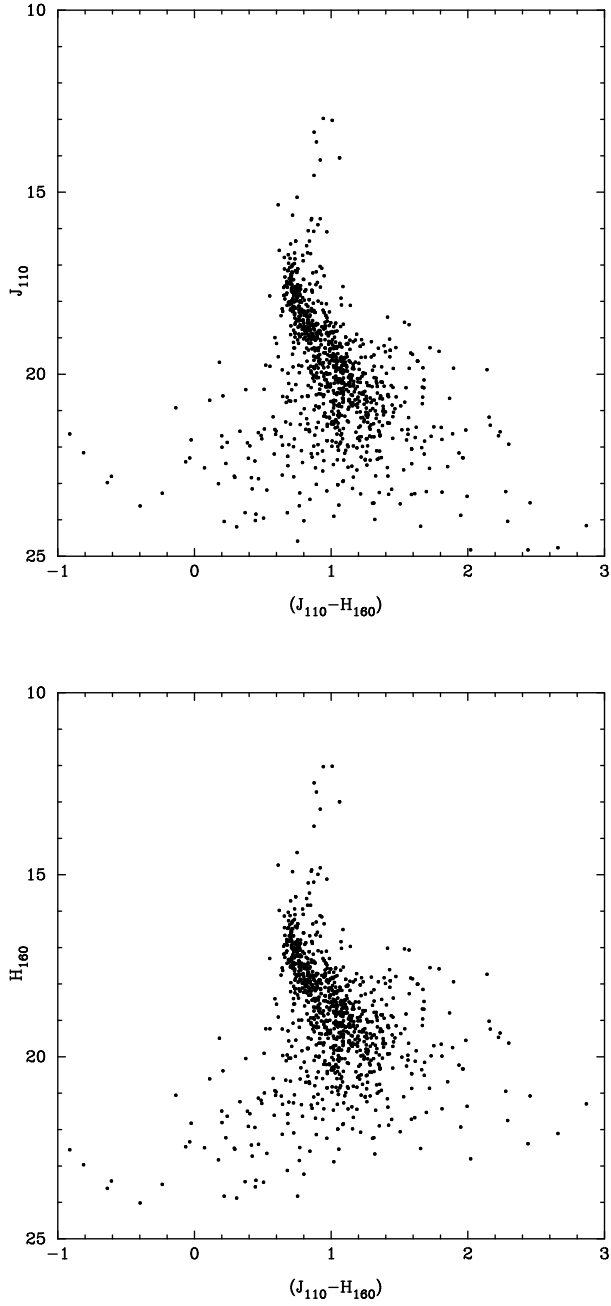}
\caption{NICMOS NIC2 $J$ and $H$ colour magnitude diagrams for NGC6553. 
The substantially reduced width of the main sequence turnoff, relative
to the optical C-M data, show that reddening is a major contributor to
the appearance of the photometry for this cluster. \label{nic-cmds}}
\end{figure}

\clearpage

\begin{figure}
\psfig{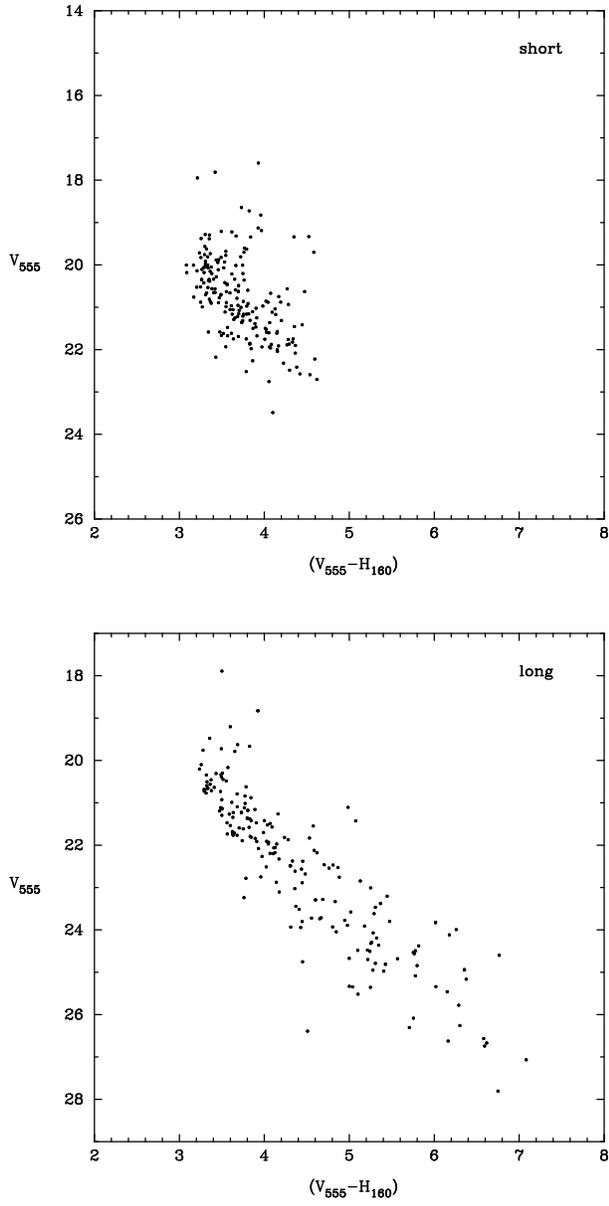}
\caption{Optical and NICMOS colour-magnitude relations for
NGC6553. The significant increase in width of the main sequence in
the colour-magnitude relations involving optical colours, compared to
that in purely infrared colours, is indicative of variable extinction
across the cluster, even on the scale of the NIC2 field of view.
\label{nic-cmds2}}
\end{figure}

\clearpage

\begin{figure}
\psfig{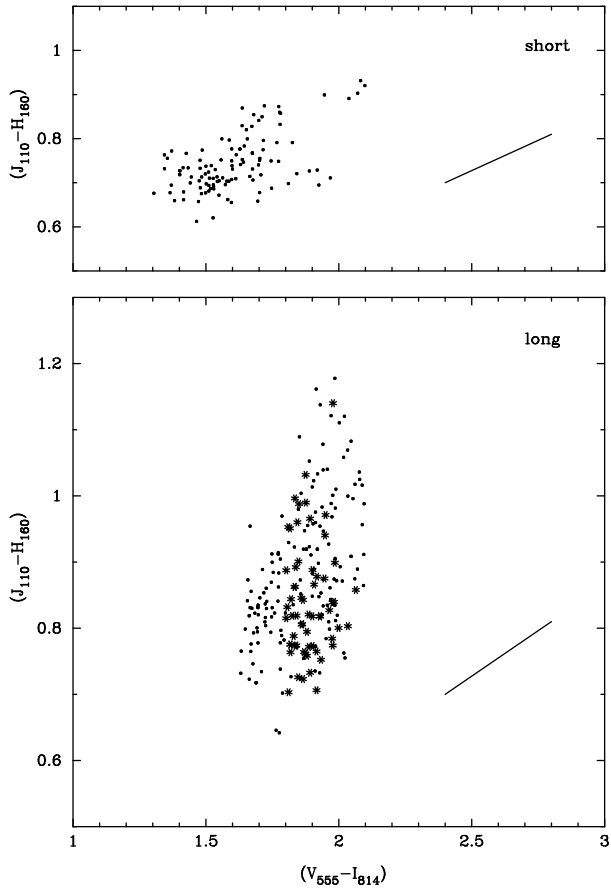}
\caption{The optical-infrared two-colour diagram for NGC6553. The
slope of the reddening vector is also shown. The top panel shows 
stars from the
short exposure WFPC2 data, with $V<21$, $V-I<2.2$, essentially the
main sequence turnoff and lower RGB. The lower panel shows data from
the long exposure WFPC2 data, with $21.0<V<23.0$, $V-I<2.2$,
essentially the main sequence just below the turnoff. Stars from the
`red flare', with $21.0<V<22.0$, $V-I>1.8$, are indicated by
open star symbols. These stars are systematically bluer in $J-H$ than
cluster stars of the same optical colour, consistent with their being
a lower metallicity background bulge population. Note also the
significant width of the colour distribution.
This illustrates that part of
NGC6553 is affected by variable extinction of amplitude some 30\% of
the total line of sight extinction towards the cluster. It is this
variable extinction which complicates analysis of the photometry for
this cluster. \label{nic-2col}}
\end{figure}

\clearpage

\begin{figure*}
\caption{ A comparison of STIS and WFPC2 magnitudes. The top left
panel compares STIS $I_{LP}$ and WFPC2 $I_{814}$ magnitudes,
stars in common. The lower left
panel compares STIS $I_{LP}$ and WFPC2 $V_{555}$ magnitudes,
stars in common. In each case there are $\sim 1600$~stars in common.
The right hand side panels show the residual colour dependence from
the STIS {\sl vs} WFPC2 calibrations. The solid lines are the fits
described in the text. SEE FIGURE 22 (JPG) \label{STIS-WFPC}}
\end{figure*}

\clearpage



\begin{deluxetable}{lll}
\tablewidth{0pt}
\tablecaption{Basic parameters for NGC6553 \label{N6553-summ}}
\tablehead{
\colhead{Parameter} &
\colhead{} & 
\colhead{Reference}}
\startdata
RA (J2000.0) & $18^{\rm{h}}~09^{\rm{m}}~17.6^{\rm{s}}$ & new measure,
this paper \nl
Dec (J2000.0) & $-25^\circ~54'~31.3\arcsec$ & \nl
$(\ell,b)$ (J2000.0) & $5^{\circ}.2529, ~-3^{\circ}.0298$ & \nl
$\rm [Fe/H]$ & $-0.4$ & Barbuy et al. (1999) \nl
$\rm [Fe/H]$ & $-0.1$ & Carretta et al. (2001) \nl
Z & $\rm \approx Z_{\odot}$ & Barbuy et al. (1999) \nl
Age & $12 \pm 2$ Gyr & Ortolani et al. (1995) \nl
$r_{c}$ & 0.55 arcmin & Harris (1996) \nl
$r_{h}$ & 1.55 arcmin & Harris (1996) \nl
$D_{\odot}$ & $\approx 4.4$ kpc & derived \nl
$E(V-I)$&0.87 & minimum, highly variable \nl
$E(B-V)$&0.72 & minimum, highly variable \nl
$(m-M)_\circ$ & 13.2 & suggested here \nl
$\rm [Fe/H]$ & $\sim0.0$ & suggested here \nl
Age & $\sim13$ Gyr & suggested here \nl
\enddata
\end{deluxetable}

\clearpage


\begin{deluxetable}{cccc}
\tablewidth{0pt}
\tablecaption{WFPC2 Datasets for NGC6553 \label{N6553-wfpc}} 
\tablehead{
\colhead{Dataset} & 
\colhead{Filter} & 
\colhead{Exposure} & 
\colhead{Date} \\
\colhead{} & 
\colhead{} & 
\colhead{(s)} & 
\colhead{}}
\startdata
u4ax1301r & F555W & 5 & 24 April 1998 \nl
u4ax1302r & F555W & 5 & 24 April 1998 \nl
u4ax1303r & F555W & 5 & 24 April 1998 \nl
u4ax1304r & F555W & 200 & 24 April 1998 \nl
u4ax1305r & F555W & 200 & 24 April 1998 \nl
u4ax1306r & F555W & 200 & 24 April 1998 \nl
u4ax1307r & F814W & 20 & 24 April 1998 \nl
u4ax1308r & F814W & 20 & 24 April 1998 \nl
u4ax1309r & F814W & 20 & 24 April 1998 \nl
u4ax130ar & F814W & 200 & 24 April 1998 \nl
u4ax130br & F814W & 200 & 24 April 1998 \nl
u4ax130cr & F814W & 200 & 24 April 1998 \nl
\enddata
\end{deluxetable}

\clearpage


\begin{deluxetable}{ccc}
\tablewidth{0pt}
\tablecaption{STIS Datasets for NGC6553 \label{STIS-data}} 
\tablehead{
\colhead{Dataset} & 
\colhead{Exposure} & 
\colhead{Date} \\
\colhead{} & 
\colhead{(s)} & 
\colhead{}} 
\startdata
o4ax14010  &  30 & 5 March 1998 \nl
o4ax14020  &  300 & 5 March 1998 \nl
o4ax14030  &  2046 & 5 March 1998 \nl
\enddata
\end{deluxetable}

\clearpage


\begin{deluxetable}{cccc}
\tablewidth{0pt}
\tablecaption{NICMOS NIC2 Datasets for NGC 6553 \label{NIC-data}} 
\tablehead{
\colhead{Dataset} & 
\colhead{Filter} & 
\colhead{Exposure} & 
\colhead{Date} \\
\colhead{} & 
\colhead{} &
\colhead{(s)} & 
\colhead{}}
\startdata
n4ax12hcq & F110W & 160 & 2 March 1998 \nl
n4ax12hiq & F110W & 514 & 2 March 1998 \nl
n4ax12h5q & F160W & 576 & 2 March 1998 \nl
n4ax12heq & F160W & 1026 & 2 March 1998 \nl
\enddata
\end{deluxetable}

\clearpage


\begin{deluxetable}{ccc}
\tablewidth{0pt}
\tablecaption{NICMOS NIC2 Photometric keywords \label{NIC-words}} 
\tablehead{
\colhead{Filter} & 
\colhead{PHOTFNU} & 
\colhead{ZPVEGA} \\
\colhead{} & 
\colhead{$\rm Jy \, DN^{-1}$} & 
\colhead{Jy}}
\startdata
F110W & 2.288E-6 & 1773.7 \nl
F160W & 2.337E-6 & 1039.3 \nl
\enddata
\end{deluxetable}

\clearpage


\begin{deluxetable}{cccc}
\tablewidth{0pt}
\tablecaption{Photometric Conversion Coefficients \label{phot-coeff}}
\tablehead{
\colhead{Passband} & 
\colhead{$(V_{555}-I_{814})$} & 
\colhead{$(V_{555}-I_{814})^2$} & 
\colhead{ZP}}
\startdata
$V_{555}$  & $-$0.5184  & $-$0.0502 & $-$23.473 \nl
$I_{814}$ & $+$0.4742  &  $-$0.0487  & $-$23.464 \nl
\enddata
\end{deluxetable}

\end{document}